%% file: main.tex
\documentclass[manuscript,natbib=false,nonacm]{acmart}

\usepackage[style=numeric,backend=bibtex,sorting=none]{biblatex}
\usepackage{makecell}
\usepackage{graphicx}
\usepackage{subcaption}
\usepackage{longtable}
\usepackage{array}
\usepackage[inline]{enumitem}  
\usepackage{amsmath}
\usepackage{pdflscape}
\AtBeginDocument{%
  }

\renewcommand\footnotetextcopyrightpermission[1]{} 
\setcopyright{none}

\begin{document}

\title{Enriching Moral Perspectives on AI: Concepts of Trust amongst Africans}

\author{Lameck Mbangula Amugongo}
\email{amugongol@gmail.com}
\orcid{0001-6468-2643}
\affiliation{%
  \institution{Namibia University of Science \& Technology}
  \streetaddress{13 Jackson Kaujeua}
  \city{Windhoek}
  \country{Namibia}
  \postcode{9000}
}

\author{Nicola J. Bidwell}
\affiliation{%
  \institution{Rhodes University}
  \streetaddress{Makhanda}
  \country{South Africa}}
  \affiliation{%
  \institution{International University of Management}
  \country{Namibia}}
    \affiliation{%
  \institution{Charles Darwin University }
  \country{Australia}}
\email{nic.bidwell@gmail.com}

\author{Joseph Mwatukange}
\affiliation{%
  \institution{Meyabase}
  \streetaddress{}
  \city{Windhoek}
  \country{Namibia}}
  \postcode{}
\email{jmwatukange9@gmail.com}

\renewcommand{\shortauthors}{Amugongo, Bidwell and Mwatukange}

\begin{abstract}
The trustworthiness of AI is considered essential to the adoption and application of AI systems. However, the meaning of trust varies across industry, research and policy spaces. Studies suggest that professionals who develop and use AI regard an AI system as trustworthy based on their personal experiences and social relations at work. Studies about trust in AI and the constructs that aim to operationalise trust in AI (e.g., consistency, reliability, explainability and accountability). However, the majority of existing studies about trust in AI are situated in Western, Educated, Industrialised, Rich and Democratic (WEIRD) societies. The few studies about trust and AI in Africa do not include the views of people who develop, study or use AI in their work. In this study, we surveyed 157 people with professional and/or educational interests in AI from 25 African countries, to explore how they conceptualised trust in AI. Most respondents had links with workshops about trust and AI in Africa in Namibia and Ghana. Respondents' educational background, transnational mobility, and country of origin influenced their concerns about AI systems. These factors also affected their levels of distrust in certain AI applications and their emphasis on specific principles designed to foster trust. Respondents often expressed that their values are guided by the communities in which they grew up and emphasised communal relations over individual freedoms. They described trust in many ways, including applying nuances of Afro-relationalism to constructs in international discourse, such as reliability and reliance. Thus, our exploratory study motivates more empirical research about the ways trust is practically enacted and experienced in African social realities of AI design, use and governance.  
\end{abstract}



\maketitle

\newpage

\section{Introduction}

Discussion about trust in AI has intensified since the European Union (EU) published its Ethics Guidelines for Trustworthy AI, \cite{EUTrustAI}. Yet, narratives about what trust actually entails in designing, using and regulating AI vary widely across industry, research and policy spaces \cite{Reinhardt23}. Professionals in industries that develop and use AI define “trustable AI-based software” according to the specific software, their personal experiences and their social relations at work, not formal definitions \cite{Pinketal2024}. However,  insights about the meaning of trust and AI are, like most research about people and AI,  situated in societies that dominate AI research and development \cite{Yu_Gupta_Rosenfeld_2023, benk2023decades, Ewuoso2023}.  These societies are Western, Educated, Industrialised, Rich and Democratic (WEIRD) \cite{septiandri2021}. Africa constitutes over a quarter of the world's countries and a third of all languages. However, so far, studies about trust and AI in Africa include just South Africa \cite{Gillespie2023} and Mozambique \cite{Beltrão23}, and do not include the meaning of trust to African people who develop, promote, study or use AI in their work. In this paper, we begin to explore how African values are engaged in conceptualising trust in AI by people with professional or educational interests in AI in Africa.

Trust is a complex and subtly varying phenomenon. Discussions about AI generally agree that trust involves a trustor who delegates a goal to a trustee and accepts the risk of incurring a loss \cite{Jacovi2021, Toreini2020, benk2023decades}. However, there are contentions about whether trust can be applied when systems are not volitional moral agents that have emotive states and can be held responsible for their actions \cite{LEWIS202233}. Concerns also pertain to how the terminology of trusted AI systems and trustworthy AI places responsibility on users, not developers or institutions \cite{Bryson} and are exploited by industry to gain end-user acceptance of data extraction \cite{Krüger23}. Older debates propose that people “confuse” trust with predictability, confidence or cooperation \cite{meyer1995}, and newer frameworks use system-like constructs (e.g. reliability, functionality, helpfulness and explainability) to differentiate from human characteristics (e.g. integrity, competence and benevolence) \cite{Lankton2015-sf, Ryan2020, Ferrario2022}. Yet, AI systems are actors in sociotechnical systems \cite{MARSH2020100039} and people do attribute intent and human-like reasoning to AI \cite{Jacovi2021} and express trust in AI according to its perceived humanness \cite{Nass94}. Thus, trust in people, communities, organisations, AI systems and data entwine and “cannot be teased apart in practice” \cite{knowles2016}.

A recent study shows that Kenyans who use  AI in software development, graphic design, journalism or advertising are concerned about intellectual property, data privacy, biased data and algorithmic limitations \cite{ngaruiya2023domestication}. Approaches to trustworthiness and trusted AI internationally, that seek to alleviate such concerns, derive constructs  (e.g., fairness, accountability, transparency) from knowledge practices, experiences and values in WEIRD societies  \cite{laufer2022, septiandri2021, Capel2023, Birhane2022}. These constructs might not be universalisable, since the concept of trust does not always easily translate between worldviews and languages \cite{Rawson2023}.  For instance, in Yoruba and Tswana (widely spoken languages in West and southern Africa) the words "gbekele" and "tshepa" are used in translating the concept of trust but actually mean, respectively, dependability and expectation that others will behave responsibly \cite{Ewuoso2023}. Indeed, studies of trust in technologies other than AI, in Africa, show that constructs about trust must account for local norms \cite{Weichselbraun2023-tv} and global relations of power, capital and geography (e.g. \cite{vanStaden2023, Mhlambi2023-rf, Birhane2022}. In the past five years, scholars have introduced Afro-relational philosophies into international conversations about the ways data governance can help balance historical power differentials \cite{Olorunju2023}, and theory in AI ethics (e.g. \cite{ Birhane2022, Mhlambi2023-rf, Okyere-Manu2023-fs, Abebe2021}). Some blend Afro-relationalism with constructs prevalent in international AI discourse \cite{Amugongo2023Ubuntu} and others contrast concepts of trust in Africa and WIERD societies \cite{Ewuoso2023}.  However, how African logic and moral values contribute to contextualising trust in AI in African contexts has not been empirically explored thus far.

We contribute important, preliminary insights about values and experiences that shape how trust in AI is interpreted in industry, research and education in Africa. We surveyed 157 people with professional and/or educational interests in AI from 25 African countries. Most respondents had links with workshops about trust and AI that fostered networks amongst Africans developing, using or studying AI \cite{TrustDLIAI23, TrustAI23}. Respondents’ educational attainment, transnational mobility or country of origin influenced their concerns about AI systems, distrust in some applications and focus on certain constructs used to operationalise trust. They often expressed that their values are guided by the communities where they grew up and emphasised communal relations over individual freedoms. While they described trust in diverse ways, many respondents applied nuances of Afro-relationalism to constructs prevalent in international AI discourse, such as reliability and reliance. Thus, our exploratory study motivates for empirical research about the ways trust is practically enacted and experienced in the social realities of AI design, use and governance in Africa.

\section{Related work}
Trust and trustworthiness in relation to AI has different meanings in government, industry, and across research in software engineering, computer science and social sciences \cite{Reinhardt23, Ahmad23}. This reflects various legal, ethical and technical emphases and, as we outline next, how meanings about trust are contextual. 

\subsection{WIERD contexts for understanding trust}
A survey across five WEIRD countries suggests that trust in AI is affected by understanding and familiarity with AI, uncertainties about societal impact, and beliefs about the adequacy of regulations and laws to make use safe \cite{Gillespie2021}. Distrust arises from corporate business goals, the pace of AI development, the potential of AI systems to outperform humans and impact labour forces, livelihoods and survival \cite{Petersen2022,kevinRoose2023}, and failures of AI systems \cite{Lohr_2021,storey2022}. Failure has many sources including data quality \cite{Chakravorti_2022}, choice of algorithm \cite{hooker2021}, language handling  \cite{Amugongo2018}, adapting models to real-world data \cite{Davenport2018}, and biases in data and decisions \cite{Jabbour2023}. Studies of people’s trust in different AI use-cases in South Africa \cite{Gillespie2023} and Face Recognition Software (FRS) in Mozambique \cite{Beltrão23} show optimism, despite recognising risks, shaped by beliefs that AI systems can mitigate severe problems in local administration and infrastructure and perceptions of AI-systems’ robustness, competence and benevolence.
 
Measures of trustworthiness and constructs used to operationalise trust in AI are derived in WEIRD societies and vary for different AI systems. For instance, reliability, transparency and competence are critical for trustworthy autonomous vehicles \cite{benk2023decades}, but healthcare professionals' trust in medical AI is also affected by explainability, interpretability, usability, and education \cite{Tucci2021}. Scholars increasingly contest ‘one size fits all’ approaches to trust \cite{Castelfranchi2016, Harper_2014, Knowles2015, Lankton2015-sf,benk2023decades}. They advocate for contextualised frameworks that account for differential trust in AI systems \cite{Gillespie2021} and how a trustor’s expectations of qualities associated with trusted AI are shaped by features of the situation \cite{Berkel2022}, past experiences in situations they consider to be similar \cite{Reynolds2010} and their professional identity \cite{Leavitt2012}. However, context and activity are inseparable and frameworks can’t delimit the dimensions of context because contextual features are always produced, maintained and enacted within particular occasions \cite{Dourish2004}.
 
Trust’s contextual nature explains why professionals in industries that develop and use AI align concepts of trust and trustworthiness to their everyday realities and sociotechnical relationships at work \cite{Pinketal2024}. Pink et al \cite{Pinketal2024} find that professionals do not refer to industry measures of trustworthiness or models to operationalise trust in AI (e.g., \cite{Amugongo2023SDLC, Lukyanenko2022-fa}. Sociality shapes people’s sense and discernment of what is acceptable or appropriate, and people's tastes and manners are embodied \cite{Garrett23} and reflect their position in social spaces \cite{Bourdieu1984}. The politics of aesthetic sensibility and discernment shapes everyday work in tech industries \cite{Seaver2022}. However, Pink et al \cite{Pinketal2024} do not say how the engineers, developers and marketing and strategy managers, that they interviewed about trust, are culturally situated, though they seem to be in WEIRD societies because they were recruited via the researchers’ contacts and almost all were given European pseudonyms.

\subsection{Epistemic norms, power and technology in Africa}
Epistemic norms \cite{Kauppinen2018} can exclude people’s accounts or undervalue or distort their meanings when their experiences, or ways of describing their experiences, are unfamiliar \cite{Fricker2017}. Thus ethical AI must start, as US-based Ethiopian Timnit Gebru urges, by considering “who is creating the technology, and who is framing the goals and values of AI” \cite{Gebru2020}. The same is true for conceptualising trust \cite{LeeRich2021}. Since daily realities and sociotechnical relationships shape trust \cite{Pinketal2024}, people with characteristics that are underrepresented in professions that study, develop or use AI (e.g. gender or race, \cite{West2019}) may be more aware of some of the risks in using AI and particular data or predictors. Consider, for instance how a Ghanaian-American woman founded the Algorithmic Justice League before racially prejudiced algorithmic assessments \cite{Mehrsbi2021, Hill_2020, obermeyer2019} or racial stereotyping in FRS \cite{Bianchi2023} were reported. Also consider how it was an African who noticed that when primed with African, but not English, names a generative AI produced images of baboons  \cite{BlackLivesMatter2022}, a representation often experienced as a racist slur \cite{Cornish2022}.

African scholars have long recognised the apparatus of race in digital technologies \cite{Mbembe2017}.  On the one hand, rhetoric about Africa’s Fourth Industrial Revolution (4IR) \cite{Chinaza2020,kaisara2021} has provoked calls for Pan-African resistance to AI on the continent \cite{Nyagadza2021, Mapuranga2021}. On the other, African professionals are pursuing trust and AI. Namibian developers, entrepreneurs and activists considered a manifesto for ‘Building Tech ‘We’ Trust‘ to foster local accountability in safety, security, privacy, responsible data collection, transparent data management and reliable services and recognise the specific needs of people without smart devices or internet connections at {‘TrustAI’} workshop \cite{TrustAI23}. Informed by interview studies Mozambican scholars suggest that AI systems should “dampen” the trust of vulnerable groups \cite{Beltrão23}. Meanwhile, at the {‘Trustworthy AI’} workshop in Ghana, experts in machine learning and AI spoke about the impact of biased training samples on African populations and alternatives, such as small data models and “Thinking of Trust beyond Data and Models” \cite{TrustDLIAI23}.

Post-colonial techno-geopolitics shape vocabularies in technology discourses in Africa, including about trust \cite{Bidwell2021}. Regulations \cite{Bentotahewa2022}, ethical codes \cite{Wong2024}, and technology methods \cite{Avle2017} and research \cite{Okolo2022, Bidwell2022-cw, laufer2022, septiandri2021} in WEIRD societies have normative power elsewhere. Consider how the webpages of the {‘Trustworthy AI’’} and {‘TrustAI’} workshops \cite{TrustDLIAI23, TrustAI23} use terms prevalent in international discourse, and used to operationalise trust: robustness, explainability, privacy, fairness, security, transparency, equitability and accessibility. Use of these terms has real impacts on African practitioners’ life worlds by conferring international legitimacy, and social and economic capital \cite{Avle2017, Wong2024}. Yet, WEIRD terms are also appropriated in local social realities. Consider how a motivation on the {‘TrustAI’} website “to grow the customer base through trust, i.e. increasing safety for both consumers and vendors” \cite{TrustAI23}, both echoes the instrumentalist use of trust as a resource by industry in WEIRD societies \cite{Krüger23} and also shows customers’ and vendor’s mutual vulnerability. 

\subsection{African ethics and relational and reliance accounts of trust in AI}
Studies about data sharing and ethnographies of trust in technology in Africa show that problems arise when assumptions about trust do not account for social realities \cite{Weichselbraun2023-tv, Birhane2022}. Trust in AI relates to values and values differ in different societies \cite{Sutrop2019}.  For instance, views in 233 countries, on the morality of decisions made by an autonomous vehicle, vary with both economic equality and individualist and collectivist values \cite{Awad2018}. Individualist approaches to ethics focus on people as individuals and are often set up in dichotomy with ethics grounded in social, political, institutional or interpersonal relationships. Afro-communitarianist ethics, for instance, are grounded in conceptualising humanness as a property that emerges in relations between people \cite{Ewuoso2023}. While relational ethics exist outside of Africa \cite{cooper2022accountability}, they did not come to Africa from elsewhere \cite{Ewuoso2023}. For instance, many African languages have a specific construct to denotes people's interdependence, communal relationships and expected conduct towards others (e.g., ubuntu, botho, obuntu-bulamu). Similar traditions and colonial experiences, and regional cooperation in Africa universalise Afro-relational values \cite{Magu2023, Uleanya} , such as group cohesion, sharing, reciprocity, inclusivity, honesty and conformity, harmony, caring, compassion and respect for the dignity of others \cite{Magu2023}.

While concerns are raised that the technologies of 4IR weaken social ties and unity \cite{Chemhuru2021, Uleanya}, empirical studies show that Afro-relationalism has roles in digital transformation \cite{Magoro2022, Bidwell2021}. Afro-relationalism has also inspired better models for datification \cite {bidwell202131}  and data sharing,  for instance, the design of Namibia’s Digital Identity system was informed by how peoples’ trust in an organisation's use of their personal data reflects that organisation's loyalty to them \cite{vanStaden2023}.  Indeed, Afro-relationalism is increasingly suggested to guide AI ethics for Africa and beyond \cite{Amugongo2023Ubuntu, Birhane2022, Mhlambi2023-rf, Okyere-Manu2023-fs, Abebe2021} often blending terms that are prevalent in international AI discourse.

Ewuoso's (2023) conceptual analysis indicates that, from an Afro-communitarianism perspective, trust is bidirectional. A trustor is optimistic that a trustee will act with goodwill and do what is expected, and a trustee shows a willingness to do what is expected. Western philosophic accounts of trust also involve reciprocal activity, goodwill and willingness, which differentiates trust from the construct of “reliance” \cite{Sutrop2019}. However, unlike Afro-relationalism, Western accounts of trust involve evidence of the trustee’s competence to do what is expected \cite{Sutrop2019, McLeod2020}. Within  interdependent relationships, however, a trustor does not need to be certain of a trustee’s abilities but only to believe a trustee will act in their best interest and try to meet expectations, based on intuiting from direct and indirect experiences \cite{Metz2022}. Further, trust is normative in Afro-relationalism and sets moral standards about regard for others in intentions and actions, which makes the trustee as vulnerable as the trustor in a trust relationship \cite{Ewuoso2023}.

\section{Methods}
To start to explore how conceptualising trust in AI may integrate values and experiences in Africa, we surveyed African people with professional and/or educational interests. Our approach was informed by prior insights about attitudes to AI trust \cite{Gillespie2021, Gillespie2023} and standards about well-defined measures \cite{McKnight2002}. Importantly, our approach was shaped by our many years of experience in teaching, research, industry and civil society discussions in Africa.

\subsection{Survey Instruments}
We arranged 36 questions in seven sets, six closed-answer and one open-ended (see Questionnaire in appendix section). Set-1 obtained demographic and geographic information and had single and multiple options. Question Set-2 explored sentiments of optimism and fear about AI; Set-3 probed different influences on respondents' values; and Set-4 examined trust in scenarios contextualising AI systems. Questions in Set-5 compared “universal” constructs used to operationalise trust in AI (consistency, reliability, explicability, accuracy, accountability, security, inclusiveness and fairness), informed by observations in southern Africa, such as civil society debates that discuss cybersecurity more than AI accountability, and research ethics protocols that require community consent prior to individual consent. Set-6 explored attitudes towards individualism and Afro-relationalism. Sets 2 to 6 used 5-point Likert scales and enabled respondents to express neutrality/uncertainty. Most scales were the extent of agreeing or disagreeing with statements such as: “There are many reasons to be worried or fearful about AI systems” (Set-2), but the scale for Set-4 was the extent of trust or distrust, such as “in an AI system used in a national hospital to treat a family member’s life-threatening disease". 

The last questions asked respondents to describe what trust means to them and whether they would like to receive information about the survey's outcome. Most respondents took 30 – 40 minutes to complete the questionnaire, displayed in a Google Form, and while not compelled to answer any question, answered them all.

\subsection{Questionnaire administration and survey sample}
We circulated a QR code, linked to the Google Form, to African people who attended the {‘TrustAI’} workshop in Windhoek, July 2023, the {‘Trustworthy AI’} workshop in Ghana, September 2023, and contacts working in several companies and institutions in Africa. We received 166 responses between July and October and removed nine responses by people who had not grown up in Africa. Thus, our sample comprised 157 people who stated they grew up in one of 25 African countries and currently work in 19 African countries and 9 countries outside of Africa. A total of 96 respondents work in their native country, 15 work in a different African country and 46 respondents work outside Africa (Fig. \ref{fig:countries}). Some 65\% were employed as trainees/early-, mid-career or senior professionals, and 18\% were students. Respondents applied their skills in many sectors and some 40\% listed research and/or education amongst their three usual occupations (Fig. \ref{fig:occupation}). As shown in Appendix: Table \ref{tab:demographics}, our sample comprised 60\% males and 40\% females (we offered five options), most aged 25 and 44 years, and nearly half aged 35 and 44 years. All respondents had completed high school and 90\% had university degrees, often to postgraduate level.

\subsection{Quantitative and qualitative analysis }
We focus the findings in this paper on the most statistically significant or qualitatively richest insights. We treated responses to all closed-answer questions quantitatively. We enumerated the Lickert scales (Strongly agree/Fully trust: 2; Agree/Trust: 1; Neither agree or disagree/distrust nor trust: 0; Disagree/Mistrust: -1; Strongly disagree/Fully distrust: -2), computed the tendencies for each question and examined differences amongst distributions of responses. We also assessed variances among responses between demographic and geographic groups. We computed t-tests to investigate variations between genders and between levels of education and compared differences between the ten countries that had more than 5 respondents. To determine statistical differences between three or more groups we used Kruskal-Wallis one-way analysis of variance, a non-parametric method for unequal sample group size \cite{Mckight2010} and used a standard p<0.05 to interpret significance.  

We quantitatively and qualitatively analysed responses to the open-ended question, "Briefly describe what trust means to you?" Six respondents did not respond to this question or stated they were unsure, 149 responses were in English (two were in Portuguese), the average response length was 16 words and the longest was 76 words. Over 135 responses were at least one complete sentence and 20 at least two. Responses comprised 2,500 words in total,  including 505 unique words/word roots. We manually computed the frequency of human- and system-like constructs used in AI trust discourse \cite{Lankton2015-sf, Ryan2020, Ferrario2022}. Then we manually open-coded responses by interpreting comments about defining trust; characters of subjects and objects; psychological and social orientations; felt experiences; intentions and effects; particular contexts and relationships. We grouped repeating codes into themes, such as “sense of safety or a lack of fear”, and responses often related to several themes.

\section{Findings}
Most respondents (119) agreed/strongly agreed that AI systems will benefit everyone in society. However, as we summarise first, trust varied with different application contexts and their educational attainment and transnational mobility (Fig. 3). We go on to show that respondents’ educational attainment also related to how they prioritised constructs, often assumed universal in operationalising trust in AI. Most respondents stated that their values were shaped by the communities in which they grew up, their education and/or the sectors in which they work. However, the relative influence of these contexts and respondents’ orientations towards individualism varied with the countries in which they grew up and their transnational mobility. We end our findings by describing how respondents conceptualise trust using system- and human-like constructs and qualities of Afro-relationalism.

\subsection{Less trust when AI involves data about people }\label{ssec:opti}

\begin{figure}[h]
    \centering
    \includegraphics[width=0.7\linewidth]{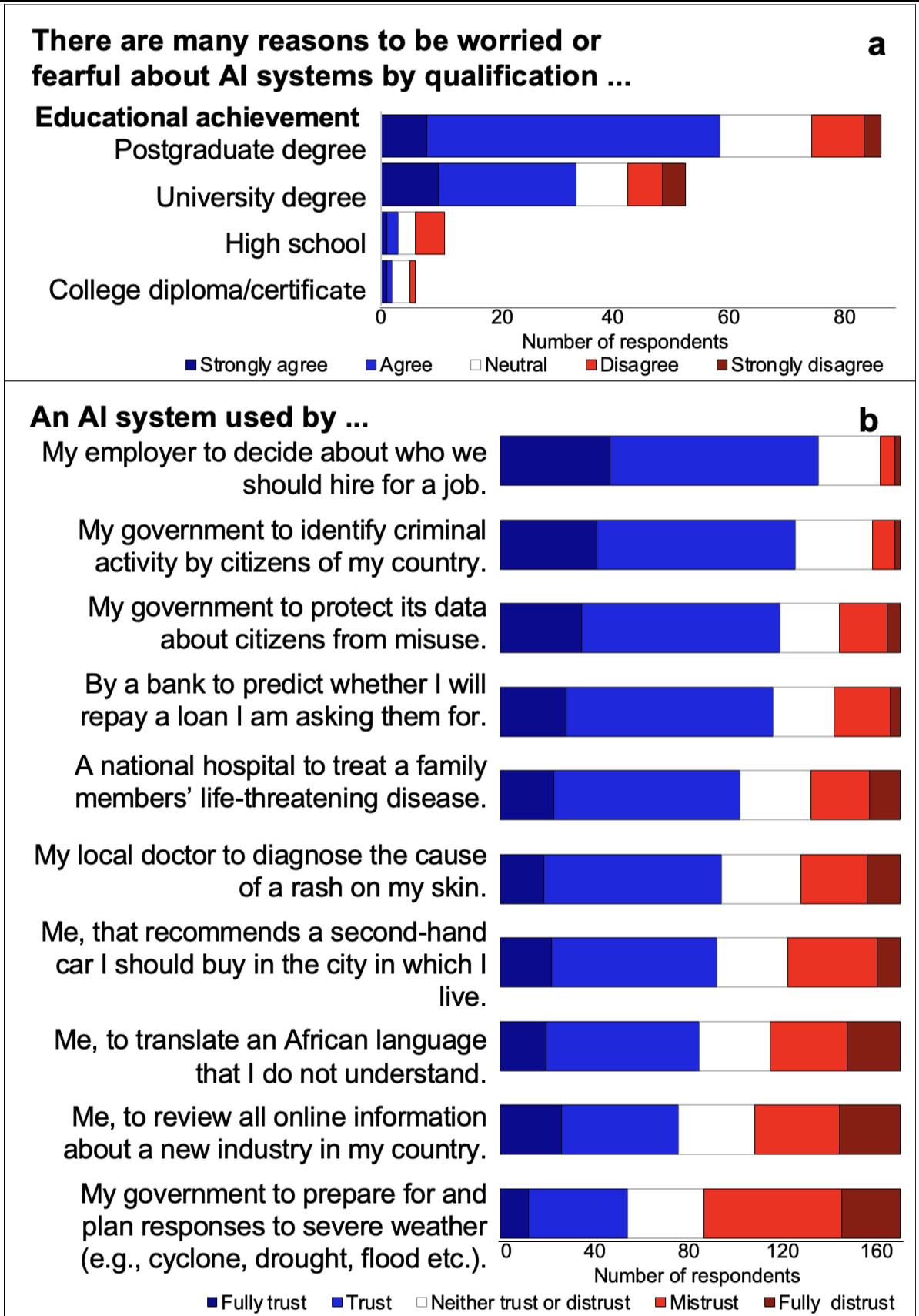}
    \caption{Responses to (a) There are many reasons to be worried or fearful about AI systems, group by respondents’ educational achievements; (b) How much would you trust an AI system in a specific scenario.}
    \label{fig:AITrustApp}
\end{figure}

Trust in AI systems depends on their applications (Fig. \ref{fig:AITrustApp} b) Respondents expressed more trust in scenarios that could, ostensibly, be limited to data about meteorology, industry, motor vehicles and languages and less trust when personal data or private information will be involved, such as for finance, government, employment and crime applications. Between 9\% and 19\% of responses across scenarios were neutral, suggesting respondents needed more information or experience to form an opinion. People who had attained higher educational qualifications were significantly more likely to agree that there are reasons to be worried or fearful about AI systems (Fig. \ref{fig:AITrustApp} a). While education, gender or location related to respondents’ trust in scenarios that were, overall, most or least trusted, there were significant differences for medical applications. While 76 of 111 respondents who live in Africa expressed trust in AI systems used “by my local doctor …” and “in a national hospital …”, only 16 of the 46 of respondents who now work outside Africa trusted AI expressed trust in these scenarios.

\subsection{Education shapes interpreting universal constructs to operationalise trust in AI}\label{ssec:prio}

\begin{figure}[h]
    \centering
    \includegraphics[width=0.7\linewidth]{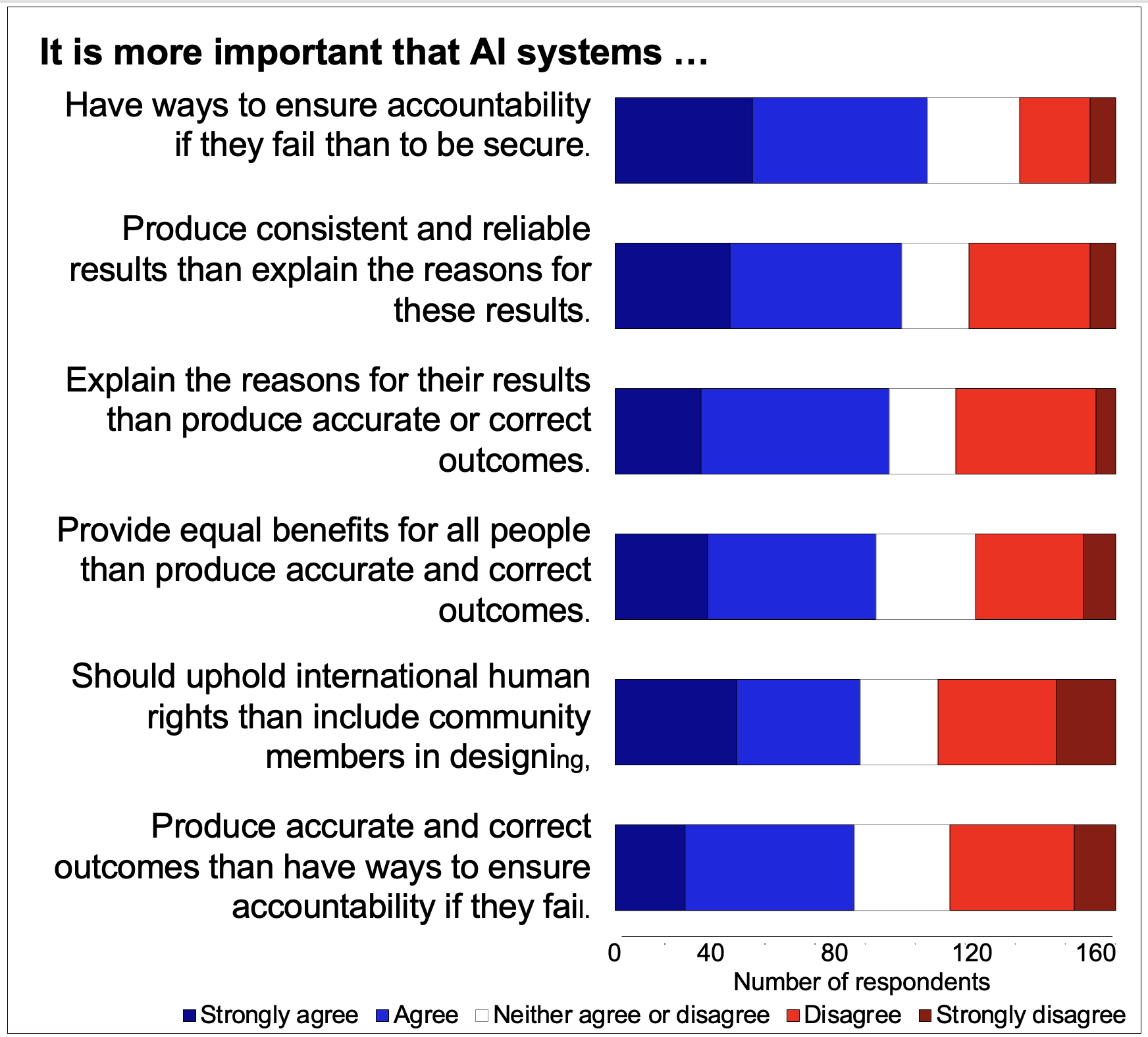}
    \caption{Responses to comparative priorities in the operation of AI systems.}
    \label{fig:AIpriorities}
\end{figure}

The relative importance of consistency, reliability, explainability, accuracy and correctness, human rights and community inclusion greatly varied to respondents, which suggests they interpreted the statements in the questions in different ways (Fig. \ref{fig:AIpriorities}). Overall, they strongly agreed that accountability is more important than security and tended to disagree that accurate results was more important than accountability (Fig. \ref{fig:AIpriorities}). However, people with postgraduate degrees (32\%) tended to disagree about prioritising accountability. 

\subsection{Influences on values and orientations towards relationalism and individualism }\label{ssec:values}

\begin{figure}[h]
    \centering
    \includegraphics[width=0.7\linewidth]{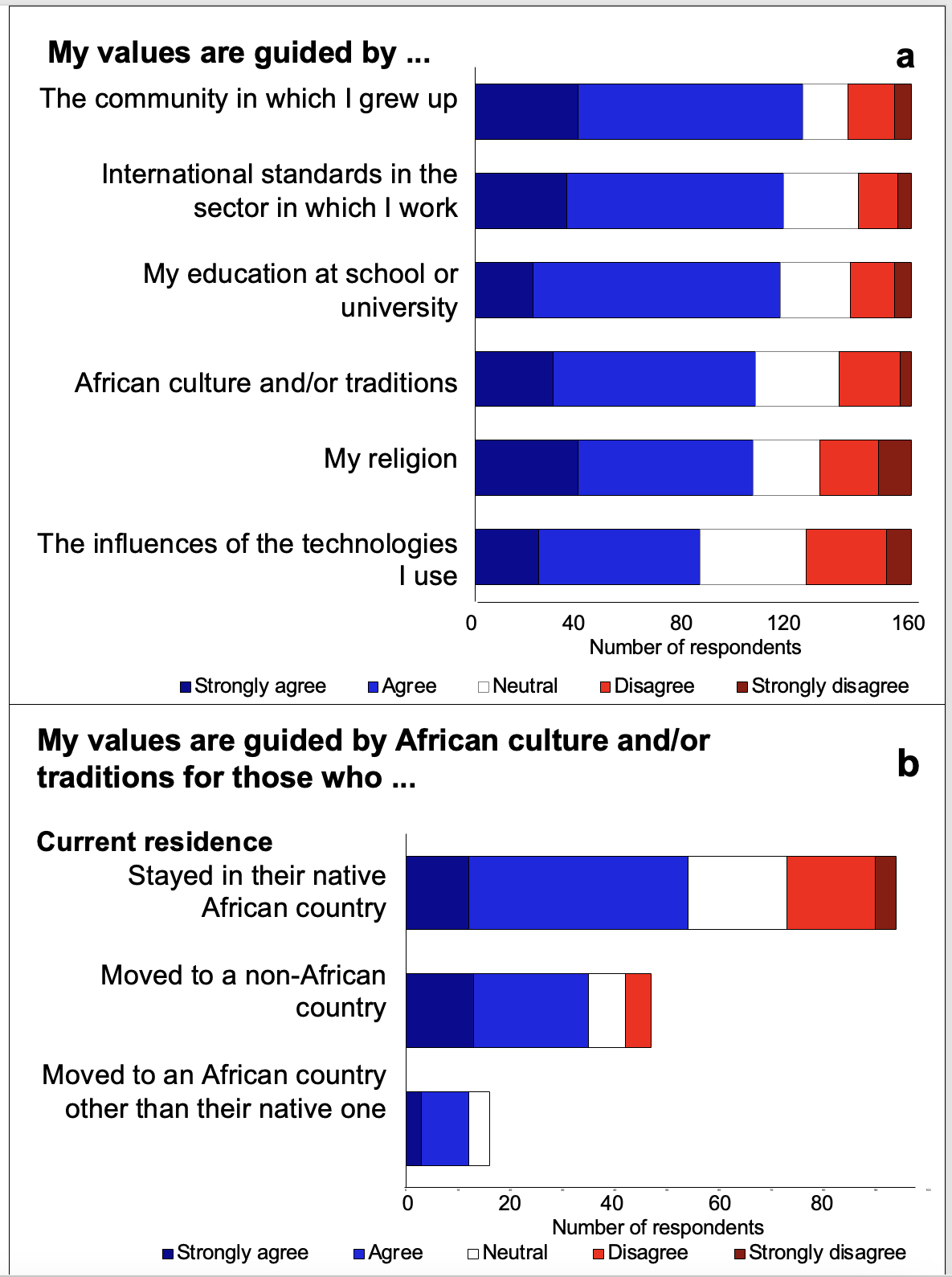}
    \caption{Responses to (a) questions about influences on values or beliefs about the right ways to think and act and (b) grouped by whether respondents now live in the country where they grew up.}
    \label{fig:AIValues}
\end{figure}

Most respondents agreed/strongly agreed that the community in which they grew up guided their values (Fig. \ref{fig:AIValues} a) and this did not significantly differ between the 10 countries where at least 5 respondents had grown up. In contrast, while overall respondents agreed/strongly agreed their values were guided by international standards of the sector in which they work and/or their formal education (Fig. \ref{fig:AIValues} a), this varied between the 10 countries where at least 5 respondents had grown up and between respondents in these countries. Compared with males, females’ values were slightly less likely to be guided by international standards at work and more likely to be influenced by religion. Responses to the influence of religion also varied across the ten countries that had at least 5 respondents and significantly differed for respondents who had relocated.

Interestingly, given their professional or educational interest in AI, respondents tended to agree their values were influenced by African culture/traditions and/or their religion more than by the technologies they use (Fig. \ref{fig:AIValues} a). Further, there is significantly stronger agreement that values are influenced by African culture/traditions amongst respondents who work in a country other than the one in which they grew up, either outside or within Africa (Fig. \ref{fig:AIValues} b).

\begin{figure}[h]
    \centering
    \includegraphics[width=0.7\linewidth]{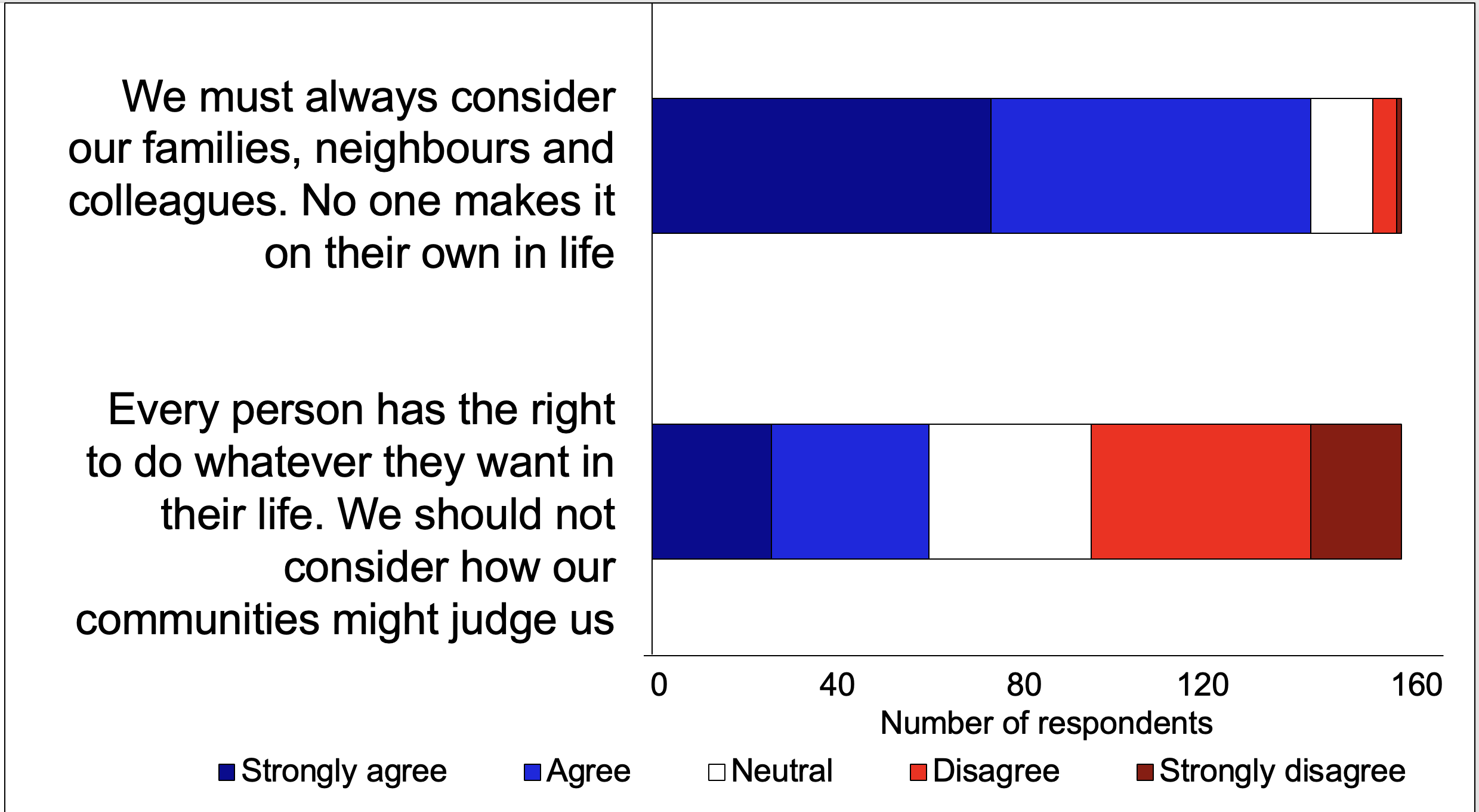}
    \caption{Responses to statements informed by Afro-relationalism (upper) and individualism (lower).}
    \label{fig:ConsiderFamilies}
\end{figure}

Respondents overwhelmingly agreed that “We must always consider our families, neighbours and colleagues. No one makes it on their own in life” (Fig. \ref{fig:ConsiderFamilies}), and those who had relocated from their native countries most strongly agreed. Responses to the contrasting statement, “Every person has the right to do whatever they want in their life. We should not consider how our communities might judge us”, however, varied (Fig. \ref{fig:ConsiderFamilies}). This variety aligns with diverse views about the relative importance of upholding international human rights and involving local community members in developing AI (Fig. \ref{fig:AITrustApp} b). Further, there is a significant difference amongst countries with at least 5 respondents (p-value = 0.039). Respondents in Namibia (55\%) and South Africa (53\%) agreed while those from Zambia (83\%), Nigeria (52\%) and Ethiopia (60\%) disagreed (Appendix: Fig. \ref{fig:everyperson}.). 

\subsection{System-and human-like qualities of trust} 

The open-ended question “Briefly describe what trust means to you” did not specify trust in AI and some responses differentiate between trusting people and things. For instance, “Believing that someone is good and honest and will not harm me, or that something is safe and reliable.” Of all responses that were complete sentences, a few relate trust to broad processes (e.g. “behaviour and actions”, “data and decision”, “process and end-result”), but most (99) frame trustees as entities. Some entities are ambiguous (5) (e.g., "something”, "object or collection of thing[s]”) but most are explicitly human or non-human (48) (e.g. “whomever or whatever”, "human or machine”, "person/organisation/AI”) or solely human (29) (e.g., "person", "individual”, "someone").

\begin{figure}[h]
\centering
\includegraphics[width=0.7\linewidth]{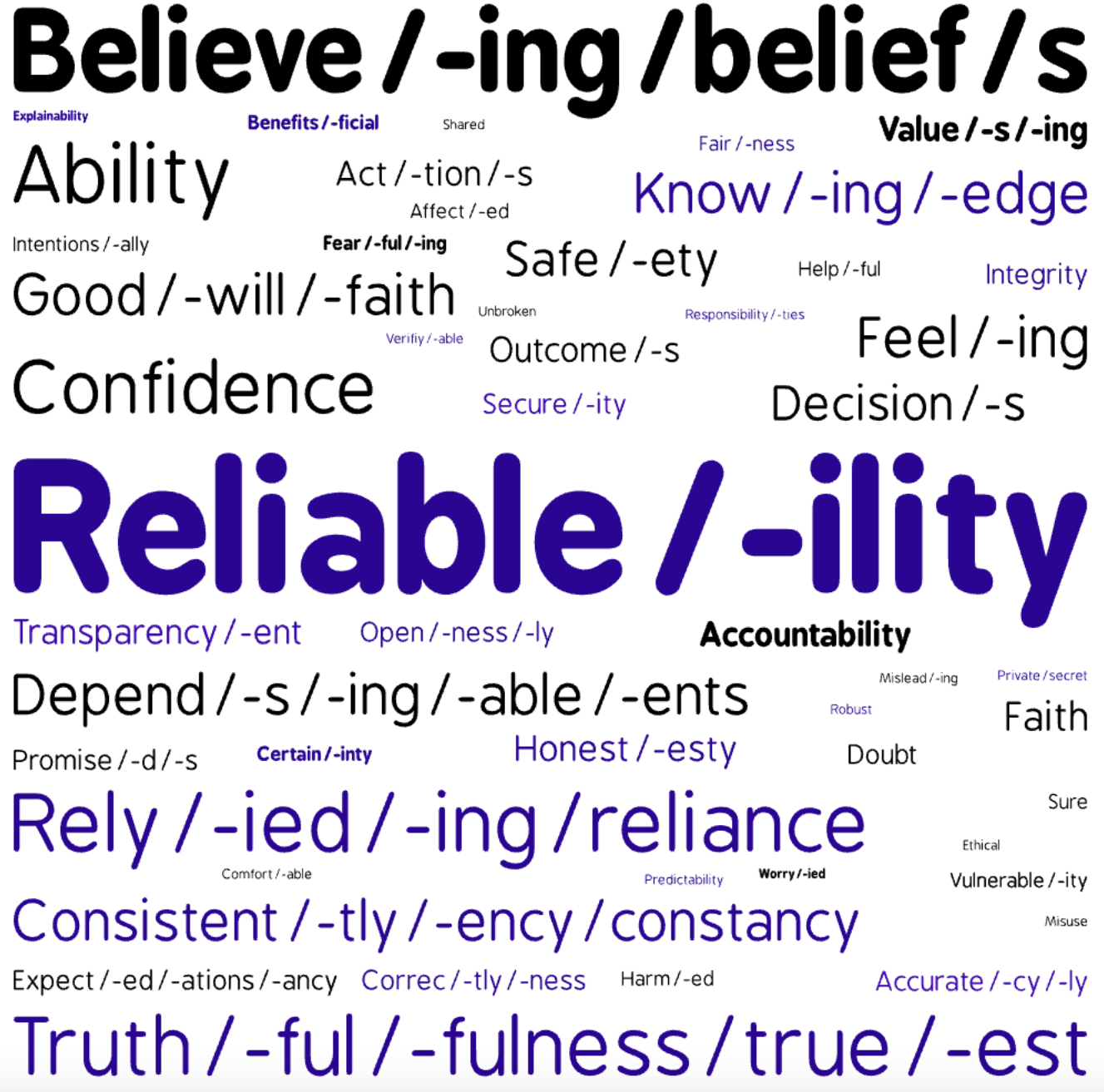}
\caption{Word frequency across all responses to the open-ended question about ‘what trust means’ to respondents. The largest font size is the highest frequency 34 instances; the smallest font size is the lowest frequency shown in this figure. Two (2) instances, purple: words common in AI discourses. Bold: words used in any section of the questionnaire.}
\label{fig:wordcloud}
\end{figure}

Many respondents used system-like constructs prevalent in XAI and AI ethics discourses (Fig. \ref{fig:wordcloud}), including words used elsewhere in the questionnaire. Responses include some constructs (e.g. reliability, reliance), others regularly (e.g. consistency, transparency, accountability), and some seldom (e.g., fairness, explainability, robustness) or not at all (e.g., interpretability, bias). The frequency of these constructs, in responses, somewhat reconciles with how respondents prioritised constructs in the closed-question \ref{ssec:prio}. However, responses tend to combine system-like with many human-like constructs that link to morality (e.g., honesty, tolerance, integrity, respect). Some responses relate trust and the ability “to accept a result, suggestion, advice, direction, etc, from a person or system without superfluous questioning” overall, however, they emphasise confidence more than certainty. Some frame links between trust and truth in an objective sense of correctness, e.g., "do not have doubts about there being errors”, but more frame links between trust and truth in subjective morality, such as “a truthful relationship with someone or something that can’t lie…” 

\subsection{Meanings about trust and Afro-relationality} 
Many responses relate trust to a risk of some loss, often explicitly and in quite extreme ways, such as “being able to turn my back to someone holding a weapon”, see Table \ref{tab:freq} in the appendix section. Responses also indicate, however, that trust is a “complex word”, “means a number of things” and trust has a multifaceted and situated character; for instance:
\begin {quote}
“For me trust depends on how vulnerable I feel at that point and which situation or circumstances I am in. It is believing that whomever or whatever I entrust with my information will do as expected and nothing more and will not use it against me in future. Trust is also affected by belief, experiences, propensity, tolerance, and fear. We trust depending on those. So these determines whether I feel safe and can trust you.”  (Female, IT Analyst in the education sector)
\end{quote}
Responses relate trust to trustees doing what was promised or expected, such as "… another person (or a software program) that we are sure will do the job for us in the best possible way”. Some expectations and acceptance refer to the “ability of someone or something”. Many, however, associate trust with intentionality that respondents contextualised in a positive value-judgement such as, “…someone or something is reliable, honest, and has good intentions” or “…inspired by being truthful, upright, dependable, reliable and having no negative intentions”. Or, in another response,
\begin{quote}
“Trust starts with truth and ends with truth but trust is not simply a matter of truthfulness or even constancy. It is also a matter of amity and goodwill.”  (Male, consultant in law sector) 
\end{quote}
Several responses associate good intentions with accountability or transparency, such as “open and honest effective communication” or knowing that if someone fails “they will openly say so”, for instance:
\begin{quote}
“Trust implies I have both capability, goodwill and fortitude to follow through on expectations and should I have to fail, provide adequate reasons and backups to ameliorate and possibly prevent damages”.  (Male, early career employee in engineering)
\end{quote}
Responses often refer to believing more than knowing. Some describe trust as a feeling or a felt experience, such as “comfort”, which may link to shared values, for instance “... is a feeling, formed by people with a common set of beliefs and values”. The agreements involved in trust can be “tacit” and may not imply in-group conformity, such as “I feel safe to be myself, like have an opinion that differs with those I’m in the company of”. Many respondents, based in Africa or not, characterised trust in Afro-relational terms, such as “respect for dignity”, for instance, 
\begin {quote}
"Having faith in someone that they'll do "their best" at something, even though the outcome may not be what we both desired. E.g. having faith that the government will do what they say, and not do the opposite of what they said." (Male, commerce student)
\end {quote}
Many responses indicate that trust involves a trustee accounting for other people’s needs, for instance, “We trust those who have our best interests at heart and mistrust those who seem deaf to our concerns”. In referring to decision-making or information privacy, several note “loyalty”, “having my back” or “without the possibility of betrayal”. Responses also indicate mutuality, for instance, “so that I raise no doubts for someone when they think of me in any aspect and vice versa”. Some responses consider trust is “the essential element that enables cooperation, openness, and a sense of security in interactions and relationships”, others emphasise that a trust is active, for instance, “Dependable. Closed eyes Fall back test”, and bidirectional, for instance,
\begin{quote}
“Trust is reciprocity, a two-way relation between a collection of things. Trust also evolves over time, it is not a one-off attribute that can be attained.”  (Male, mid-career technology researcher in the education sector)
\end{quote}
Indeed, several responses relate trust to temporality. Some through routine, such as “habit about my environment”, or “earned, over time” such as “through consistent actions”. Others consider trust requires time to be invested, for instance "… you have taken time to understand me and my values in life and that I should do the same to you.”

\section{Discussion}
Africans with professional and/or educational interests in AI describe trust in many ways. This includes linking system-like constructs, used to operationalise trust in AI (e.g., reliability, transparency), to moral judgements and referring to interdependence and moral standards of conduct towards others. Obviously, links to morality are not unique to African or relational definitions of trust and are considered in models that formalise trust in AI-derived in WEIRD societies (e.g., \cite{Jacovi2021}). However, the values of most respondents in our sample are influenced by the African communities in which they were raised, as much or more than their education or professional sectors, and by African culture/traditions and/or their religion more than the technologies they use. Responses in our survey align with Ewuoso’s (2023) Afro-relational treatment of trust, such as about shared values, truthfulness, reciprocity and vulnerability between trustee and trustor. From an Afro-relational perspective, moral standards enable a trustor to believe a trustee is benevolent and has integrity without certainty in a trustee’s capability \cite{Ewuoso2023}, which differs from how trust or reliance is usually conceptualised for AI.

Based on studying perceptions of FRS systems in Mozambique, Beltrão et al.\cite{Beltrão23} call for informing minorities in Africa about AI’s lack of neutrality. Respondents in our sample expressed distrust in AI application scenarios that were likely to use personal data in decisions by government, employers and banks. However, their definitions of trust did not mention bias and emphasised privacy and security, more than fairness. The intensity of public debate about the risks of AI affects trust in AI systems in WEIRD societies \cite{Gillespie2023} and civil society discussion in Africa has focused on information privacy and protection, prompted by consultations about new regulations about data sharing and use \cite{vanStaden2023}. Different awareness of bias might explain why respondents who work outside Africa have less trust in AI systems in healthcare given links between the use of data in medicine and racial discrimination and deception in black minorities overseas {\cite{LeeRich2021, Harrington23}. We suggest that discussions about biases in data, models and AI use must account for local nuances in trust. For instance, greater trust in an AI system used by a “local doctor” for respondents in their native country might imply a practitioner with whom they have a communal and/or long-term connection. Certainly, rural Namibians trust decisions about technologies made by community members they consider more technically skilled than themselves \cite{Aludhilu2018}. A willingness to voice concerns, however, can be stifled by communal values for the unit and by self-perceptions of their own status \cite{Widder2023}. Thus professionals in building and using AI systems must not just maintain relationships with communities and involve them in decisions at {\cite{Beltrão23, Amugongo2023Ubuntu, OLeary2020} but also carefully reflect on the ways trust is enacted in everyday relationships with people, and between people and technologies.

A few respondents mentioned felt experiences of trust, such as a sense of safety or comfort, or that people with shared values and beliefs form feelings of trust. Analyses of trust in automated systems in WEIRD societies refer to affective sensory experiences \cite{Pinketal2024, Harrington23} but have not considered how these feelings entwine with the social formation of values. The diverse ways that responses combine logic about communality, interdependence and individual freedoms suggests that meanings about trust are shaped by the different social milieus respondents encounter. People don’t always recognise how values shape affective sensory experience and some respondents note trust links to habit and tacit agreements. Respondents who relocated from their native country more often recognised that African culture/traditions shaped their values, possibly because their critical distance or ‘defamiliarisation’ makes the values that are embedded in everyday routines more visible \cite{Bell}. Approaches to “felt-ethics” in HCI advocate for cultivating awareness of how values are embodied and felt in everyday practices \cite{Garrett23}. We propose that communicating about how the felt experience of trust relates to social groups can help tackle ambiguities in narratives about trust in AI \cite{Reinhardt23}. We also propose unpacking the felt experience of trust can contribute to determining what trust entails in developing, deploying, using and governing AI in different societies.  

\subsection{Limitations and future work}
We seek to prompt further studies about the ways practitioners who develop or use AI in Africa combine local values and experiences with international meanings of trust.  This requires enabling people to articulate different ways of knowing about trust, including reason, senses, emotion, faith and intuition, on their own terms. Questionnaires are weak instruments for gaining insights into many ways of knowing and this limitation is conflated with asking questions in English and mainly recruiting respondents linked to two events. Our sample of people from just over half of Africa's countries cannot represent the diversity of a continent that has at least 75 languages with over a million speakers. English is widely used in education and business, however a quarter of Africa has less proficiency, especially in Francophone and Arabic countries.

Our study suggests that people combine logic about communality, interdependence and individual freedom in different ways, which shapes how they conceive trust for different AI applications. Recognition that trust evolves is coherent with views that values are situated in unfolding situations and trust emerges in life’s “ongoingness” \cite{Pinketal2024}. Short, written scenarios cannot depict this temporality, nor can people accurately anticipate their responses ex vivo}. Thus, we advocate for studies, in home languages, of practitioners' experiences of trust in their everyday lives and how they enact values in their AI practices.

\section{Conclusion}
Meanings about trust expressed by African people with professional or educational interests in AI can sensitize us to the diverse and nuanced ways that constructs that are prevalent in international AI discourse are translated through local logic in local contexts. Educational attainment and/or transnational mobility shaped respondents’ concerns about AI systems, distrust of some applications and prioritisation of constructs used in operationalising trust in AI. However, definitions of trust relate to values that emerge in different social realities. Responses in our survey, for instance, suggest reliability may be interpreted in a context of interdependent relations between people and be imbued with bidirectional qualities that are not assumed by the construct of reliance in international AI discourse. Values associated with Afro-relational moralities shape key elements in trust relationships, such as a trustor’s belief and confidence and a trustee’s goodwill and integrity. Thus, our exploratory study motivates for more detailed empirical research about the everyday ways that African professionals, who develop, deploy, use and regulate AI, enact trust in their relationships with others and with AI systems and their felt experiences in trusting interactions. We propose that insights from African societies can reveal what is hidden by the universal constructs used to operationalise trust and contribute to clarifying meanings across industry, research and policy narratives \cite{Reinhardt23}.

\section{Acknowledgments}
“If you want to go fast, go alone, if you want to go far, go together”: we would like to thank all the participants who made time to provide their insightful responses to our survey. 

\pagebreak
\printbibliography

\onecolumn 
\pagebreak
\appendix
\input{appendix}

\end{document}

%% file: appendix.tex
\subsection{Survey Questionnaire}\label{ssec:survey}

\begin{enumerate}
\item Your age? 
\begin{itemize}
       \item  24 or younger \item  25-34 \item  35-44 \item  45-54 \item  55 or older.
\end{itemize}

\item Your sex? 
\begin{itemize}
    \item Female \item Male.
\end{itemize}

\item Which best describes your employment? (tick one)
\begin{itemize}
       \item Director
       \item Manager
       \item Early career professional
       \item Entrepreneur
       \item Consultant
       \item Student
       \item Looking for a job.
\end{itemize}

\item What is your usual occupation? (pick up to two)
\begin{itemize}
       \item Educator
       \item Researcher
       \item Engineer
       \item IT Analyst
       \item Data Scientist.
\end{itemize}

\item In which sector do you apply your skills? (tick as many as apply)
\begin{itemize}
    \item Agriculture
    \item Biotechnology or chemical industries
    \item Data Science or Information systems analysis
    \item Education
    \item Energy
    \item Finance
    \item Fisheries
    \item Food services
    \item Government/public administration
    \item Law
    \item Manufacturing
    \item Marketing
    \item Media
    \item Medicine/health care
    \item Mining
    \item Non-profit
    \item Retail
    \item Security
    \item Technology (general)
    \item Telecommunications
    \item Transport
\end{itemize}

\item What is the highest level of education you have completed?
\begin{itemize}
       \item Less than high school \item High school
       \item College diploma / Certificate \item University degree \item Postgraduate degree.
\end{itemize}

\item In which country did you grow up?
\begin{itemize}
       \item List all countries
\end{itemize}

\item In which country/s do you currently work?
\begin{itemize}
       \item List all countries
\end{itemize}

\item Check the circle that shows how much you agree with each statement about the increasing role of AI systems in society? \\
\textbf{I feel positive that AI systems will greatly benefit everyone in society.} 

\begin{center}
\begin{tabular}{|c|c|c|c|c|}
\hline
Strongly agree & Agree & Neither agree nor disagree & Disagree & Strongly disagree \\
\hline
\end{tabular}
\end{center}

\textbf{There are many reasons to be worried or fearful about AI systems.}

\begin{center}
\begin{tabular}{|c|c|c|c|c|}
\hline
Strongly agree & Agree & Neither agree nor disagree & Disagree & Strongly disagree \\
\hline
\end{tabular}
\end{center}

\item Check the circles to show how much you agree with each statement about your values or your beliefs about the right ways to think and act.

\textbf{My values are greatly guided by International standards in the sector in which I work.} 

\begin{center}
\begin{tabular}{|c|c|c|c|c|}
\hline
Strongly agree & Agree & Neither agree nor disagree & Disagree & Strongly disagree \\
\hline
\end{tabular}
\end{center}

\textbf{My values are greatly guided by my education at school or university.} 

\begin{center}
\begin{tabular}{|c|c|c|c|c|}
\hline
Strongly agree & Agree & Neither agree nor disagree & Disagree & Strongly disagree \\
\hline
\end{tabular}
\end{center}

\textbf{My values and beliefs are greatly guided by the community in which I grew up.} 

\begin{center}
\begin{tabular}{|c|c|c|c|c|}
\hline
Strongly agree & Agree & Neither agree nor disagree & Disagree & Strongly disagree \\
\hline
\end{tabular}
\end{center}

\textbf{My values are greatly guided by the African culture or traditions.} 

\begin{center}
\begin{tabular}{|c|c|c|c|c|}
\hline
Strongly agree & Agree & Neither agree nor disagree & Disagree & Strongly disagree \\
\hline
\end{tabular}
\end{center}

\textbf{My values and beliefs are greatly guided by my religion.} 

\begin{center}
\begin{tabular}{|c|c|c|c|c|}
\hline
Strongly agree & Agree & Neither agree nor disagree & Disagree & Strongly disagree \\
\hline
\end{tabular}
\end{center}

\textbf{My values and beliefs have been greatly influenced by the technologies I use.}

\begin{center}
\begin{tabular}{|c|c|c|c|c|}
\hline
Strongly agree & Agree & Neither agree nor disagree & Disagree & Strongly disagree \\
\hline
\end{tabular}
\end{center}

\item  Check the circle that shows how much you would trust an AI system in each context. 
\textbf{An AI system used by a bank to predict how long I will live when deciding about giving me a loan.} 

\begin{center}
\begin{tabular}{|c|c|c|c|c|}
\hline
Fully trust & Trust & Neither trust nor distrust & Slightly distrust &  Fully distrust \\
\hline
\end{tabular}
\end{center}

\textbf{An AI system used by my local doctor to diagnose the cause of a rash on my skin.} 

\begin{center}
\begin{tabular}{|c|c|c|c|c|}
\hline
Fully trust & Trust & Neither trust nor distrust & Slightly distrust &  Fully distrust \\
\hline
\end{tabular}
\end{center}

\textbf{An AI system used by a national hospital to help treat a close family members’ life-threatening disease.} 

\begin{center}
\begin{tabular}{|c|c|c|c|c|}
\hline
Fully trust & Trust & Neither trust nor distrust & Slightly distrust &  Fully distrust \\
\hline
\end{tabular}
\end{center}

\textbf{An AI system used by my government to protect the data it stores about citizens from misuse.} 

\begin{center}
\begin{tabular}{|c|c|c|c|c|}
\hline
Fully trust & Trust & Neither trust nor distrust & Slightly distrust &  Fully distrust \\
\hline
\end{tabular}
\end{center}

\textbf{An AI system used by my government to identify criminal behaviour amongst citizens of my country.} 

\begin{center}
\begin{tabular}{|c|c|c|c|c|}
\hline
Fully trust & Trust & Neither trust nor distrust & Slightly distrust &  Fully distrust \\
\hline
\end{tabular}
\end{center}

\textbf{An AI system used by my employer to help decide about who we should hire for a job.}

\begin{center}
\begin{tabular}{|c|c|c|c|c|}
\hline
Fully trust & Trust & Neither trust nor distrust & Slightly distrust &  Fully distrust \\
\hline
\end{tabular}
\end{center}

\textbf{An AI system that recommends which second-hand car I should buy.} 

\begin{center}
\begin{tabular}{|c|c|c|c|c|}
\hline
Fully trust & Trust & Neither trust nor distrust & Slightly distrust &  Fully distrust \\
\hline
\end{tabular}
\end{center}

\textbf{An AI system used by my government to prepare for and plan emergency responses before a severe weather event (e.g., cyclone, drought, flood etc.)} 

\begin{center}
\begin{tabular}{|c|c|c|c|c|}
\hline
Fully trust & Trust & Neither trust nor distrust & Slightly distrust &  Fully distrust \\
\hline
\end{tabular}
\end{center}

\textbf{An AI system that I use to summarise all available information about a new industry in my country.} 

\begin{center}
\begin{tabular}{|c|c|c|c|c|}
\hline
Fully trust & Trust & Neither trust nor distrust & Slightly distrust &  Fully distrust \\
\hline
\end{tabular}
\end{center}

\textbf{An AI system that I use to translate my words into an African language that I cannot understand.} 

\begin{center}
\begin{tabular}{|c|c|c|c|c|}
\hline
Fully trust & Trust & Neither trust nor distrust & Slightly distrust &  Fully distrust \\
\hline
\end{tabular}
\end{center}

\item  Check the circles to show how much you agree with the following statements about the increasing role of AI systems in society. 
\textbf{It is more important that AI systems produces reliable results than explain the reasons for these results.} 

\begin{center}
\begin{tabular}{|c|c|c|c|c|}
\hline
Strongly agree & Agree & Neither agree nor disagree & Disagree & Strongly disagree \\
\hline
\end{tabular}
\end{center}

\textbf{It is more important that AI systems should uphold international human rights than involve all stakeholders in their development.} 

\begin{center}
\begin{tabular}{|c|c|c|c|c|}
\hline
Strongly agree & Agree & Neither agree nor disagree & Disagree & Strongly disagree \\
\hline
\end{tabular}
\end{center}

\textbf{It is more important that AI systems provide equal benefits to all people than produce accurate outcomes.} 

\begin{center}
\begin{tabular}{|c|c|c|c|c|}
\hline
Strongly agree & Agree & Neither agree nor disagree & Disagree & Strongly disagree \\
\hline
\end{tabular}
\end{center}

\textbf{It is more important that AI systems are safe and secure than have mechanism for accountability if they fail.} 

\begin{center}
\begin{tabular}{|c|c|c|c|c|}
\hline
Strongly agree & Agree & Neither agree nor disagree & Disagree & Strongly disagree \\
\hline
\end{tabular}
\end{center}

\item Check the circle that shows how much you agree with each statement about individualism and collectivism in society.

\textbf{No-one makes it on their own in life. We must always consider our families, neighbours and colleagues.} 
\begin{center}
\begin{tabular}{|c|c|c|c|c|}
\hline
Strongly agree & Agree & Neither agree nor disagree & Disagree & Strongly disagree \\
\hline
\end{tabular}
\end{center}

\textbf{Every person has the right to do whatever they want in their life. We should not have to consider how we are judged by others.}
\begin{center}
\begin{tabular}{|c|c|c|c|c|}
\hline
Strongly agree & Agree & Neither agree nor disagree & Disagree & Strongly disagree \\
\hline
\end{tabular}
\end{center}

\item Briefly describe what does Trust mean to you? 
(maximum two sentences) \\
..............................................................................................................................................................................................\\..............................................................................................................................................................................................
\item If you would you like us to receive information about outcome of this survey please enter your email address. Your data will be stored separately from the survey to ensure your anonymity. \\
..............................................................................................................................................................................................
\end{enumerate}

\newpage
\subsection{Survey demographices}\label{ssec:demogr}

\begin{longtable}[c]{c c c c c c c c}
 \caption[Demographics]{Demographics information of participants of the study. Respondents were asked to choose up to three options for their typical occupation, whereas, for the question regarding the sectors in which they apply their skills, they were encouraged to select as many as they desired. \label{tab:demographics}} \\

\hline
& & \makecell{N=157} \\
\hline

Sex & \makecell{Female \\ Male} & 
 \makecell{63 (41.1\%) \\ 94 (59.9\%)} \\ 
\hline

Age & \makecell{24 or younger \\ 25 - 34  \\ 35-44  \\ 45 - 54 \\ 55 or older} &
 \makecell{20 (12.7\%) \\ 74 (47.1\%) \\ 50 (31.8\%) \\ 10 (6.4\%) \\ 3 (1.9\%)} \\
 \hline 
Level of education  & 
\makecell{> High school \\ High school \\ College diploma/certificate  \\ University degree \\ Postgraduate degree} &
\makecell{0 \\ 11 (7\%) \\ 6 (3.8\%) \\ 53 (33.8\%) \\ 87 (55.4\%)} \\
 \hline 
Type of employment   & 
\makecell{Students \\ Looking for job \\ Trainee or early career  \\ Entrepreneur \\ Mid-career employee \\ Consultant \\ Senior employee} &
\makecell{28 (17.8\%) \\ 5 (3.2\%) \\ 25 (15.9\%) \\ 10 (6.4\%) \\ 38 (24.2\%) \\ 12 (7.6\%) \\ 39 (24.8\%)} \\ 
 \hline
 Usual occupation & \makecell{Engineer \\ IT Analyst \\ Data Scientist \\ Educator \\ Researcher \\ Student \\ Director (Executive) \\ Senior Manager \\ Middle manager \\ Manager} & \makecell{34 (21.7\%) \\ 27 (17.2\%) \\ 34 (21.7\%) \\ 45 (28.7\%) \\ 63 (40.1\%) \\ 42 (26.8\%) \\ 11 (7\%) \\ 7 (4.5\%) \\ 9 (5.7\%) \\ 17 (10.8\%)} \\
 \hline

 In which sectors do you apply your skills & \makecell{Agriculture \\ Commerce \\ Construction \\ Biotechnology or chemical services \\ Education \\ Energy or power \\ Finance \\ Fisheries \\ Food production or services \\ Government \\ Health care or medicine \\ Law \\ Manufacturing \\ Marketing \\ Media \\ Mining \\ Non-profit \\ Public administration \\ Public services (e.g., police) \\ Retail \\ Security \\ Shipping \\ Technology (general) \\ Telecommunications \\ Transport} & \makecell{16 (10.2\%) \\ 12 (7.6\%) \\ 3 (1.9\%) \\ 4 (2.5\%) \\ 66 (42\%) \\ 10 (6.4\%) \\ 21 (13.4\%) \\ 0 (0\%) \\ 5 (3.2\%) \\ 18 (11.5\%) \\ 30 (19.1\%) \\ 7 (4.5\%) \\ 5 (3.2\%) \\ 12 (7.6\%) \\ 11 (7\%) \\ 3 (1.9\%) \\ 24 (15.3\%) \\ 11 (7\%) \\ 4 (2.5\%) \\ 7 (4.5\%) \\ 6 (3.8\%) \\ 1 (0.6\%) \\ 80 (51\%) \\ 18 (11.5\%) \\ 5 (3.2\%)} \\

 \hline
 

\end{longtable}

\begin{figure}[h]
    \centering
    \includegraphics[width=0.7\linewidth]{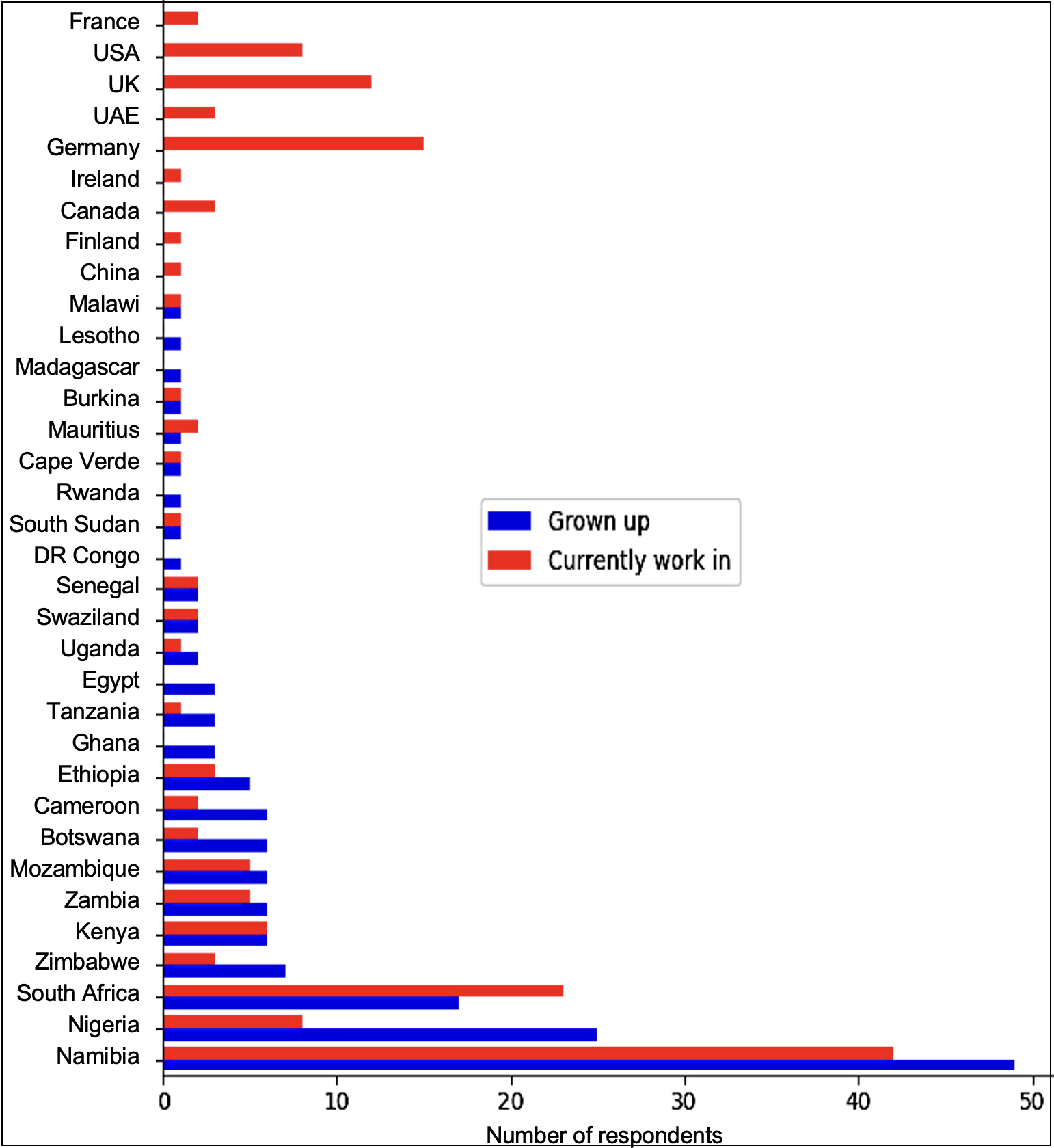}
        \caption{Countries where respondents grew up (blue) and where they currently live and work (red).}
    \label{fig:countries}
\end{figure}

\begin{figure}
    \centering
    \includegraphics[width=0.7\linewidth]{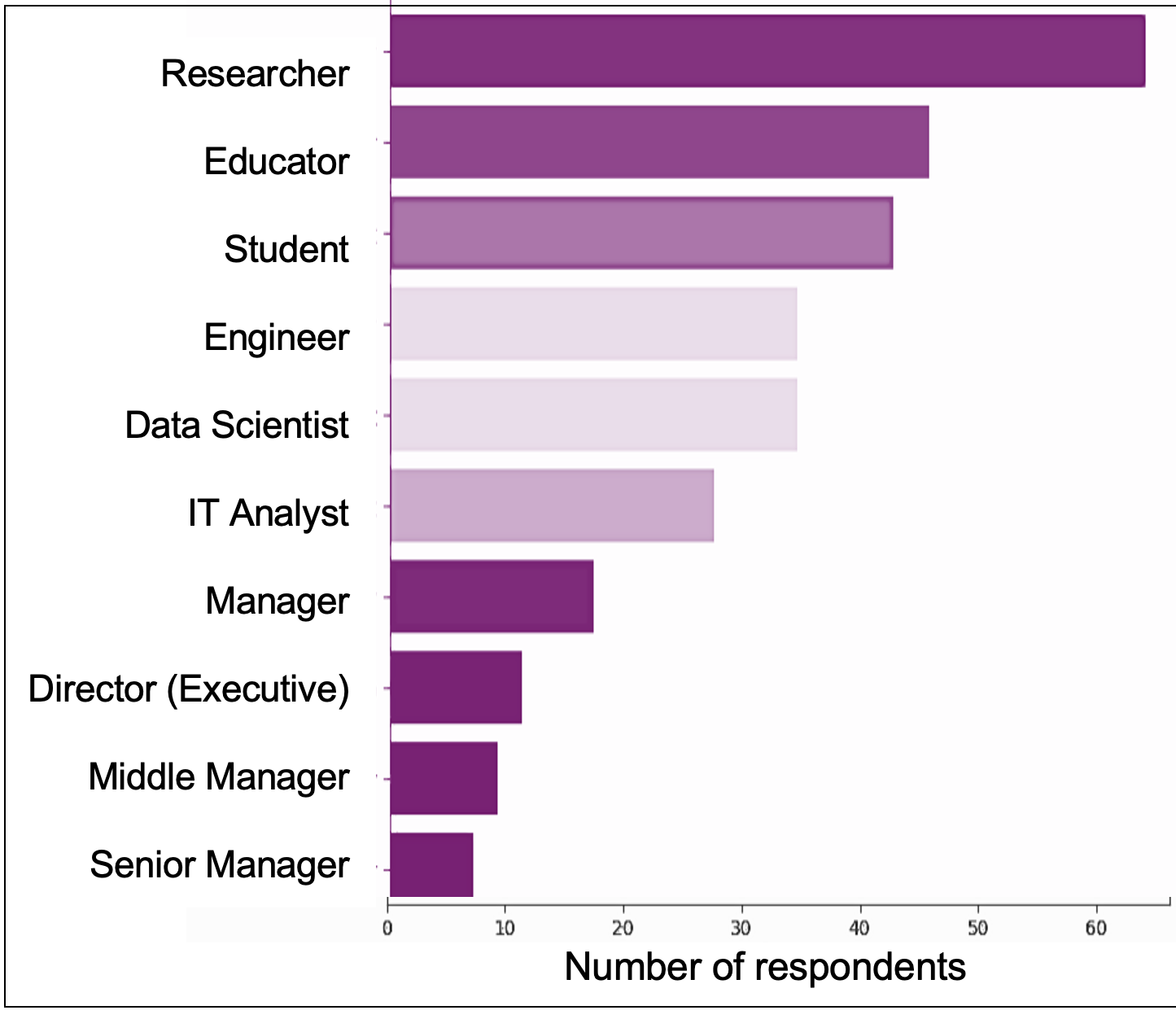}
    \caption{Respondents' usual occupations (three chosen).}
    \label{fig:occupation}
\end{figure}

\begin{figure}[h]
  \includegraphics[width=\textwidth]{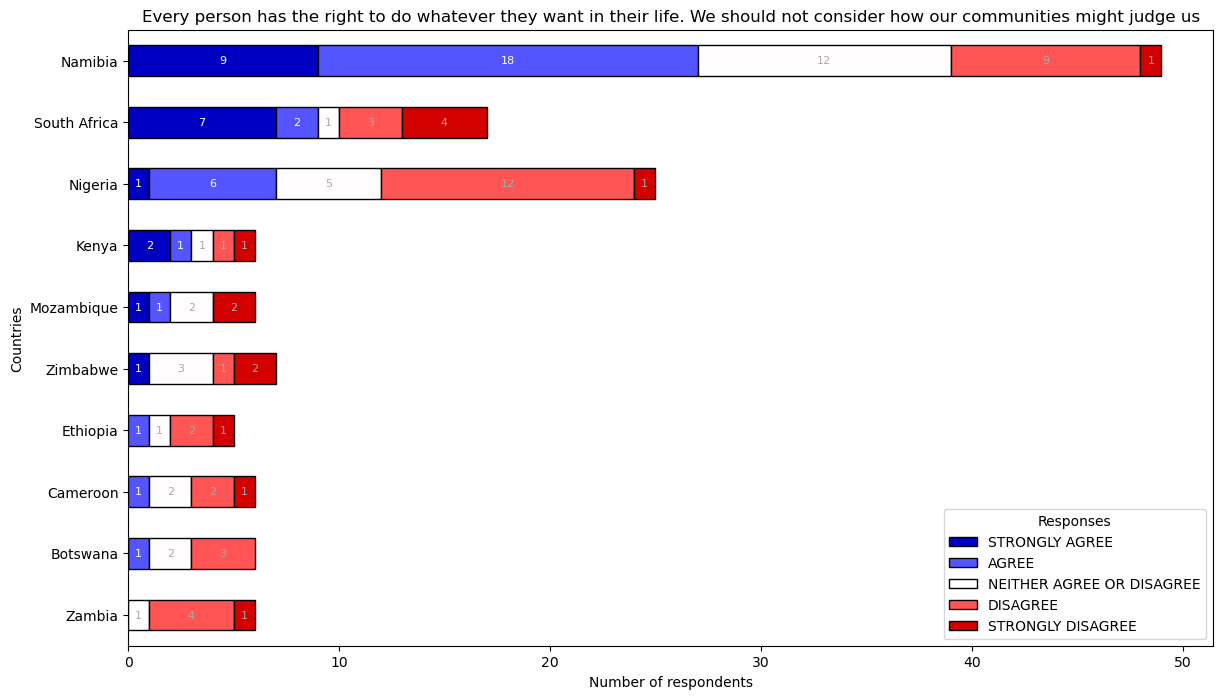}
  \caption{Responses to "Every person has the right to do whatever they want in their life. We should not consider how our communities might judge us." grouped by respondents' country where they grew up.}\label{fig:everyperson}
\end{figure}

\begin{table}[h]
    \centering
    \caption{Word frequency across all responses to the open-ended question about ‘what trust means’ to respondents. Words that are common in AI discourses are in bold; words used in any section of the questionnaire are in italics.}\label{tab:freq}
    \begin{tabular}{p{2.5cm}|c|p{2.5cm}|c}
    \toprule

        Reliable/-ility & 34 & Accurate/-cy/-ly & 5 \\
        Believe/-ing, belief/s & 28 & Integrity & 5 \\ 
        Ability & 21 & Help/-ful  & 4 \\
        Rely/-ied/-ing, reliance & 20 & Fear/-ful/-ing  & 4 \\
        Truth/-ful/-ness, true/-est & 20 & Sure & 4   \\
        Confidence  & 18 & Intentions, intentionally & 4 \\
        Depend/-s/-ing/-able/-ents & 12 & Fair/-ness & 4 \\
        Good, goodwill, good faith  & 11 & Harm/-ed &  3\\
        Consistent/-tly/-ency/constancy & 11 & Affect/-ed & 3 \\
        Know/-ing/-edge  & 11 & Vulnerable/-ity & 3 \\
        Feel/feeling  & 11 & Certain/-inty & 3  \\
        Safe/-ety  & 10 & Benefits, beneficial & 3 \\
        Decision/-s, & 9 & Comfort/-able & 2 \\
        Act/action/actions  & 8 & Shared & 2  \\
        Outcome/-s  & 8 & Worry, worried & 2  \\ 
        Transparency/-rent & 8 & Unbroken & 2  \\
        Faith &  7  & Ethical & 2 \\
        Honest/-esty & 7 & Misuse & 2 \\
        Value/-s/-ing & 7 & Mislead/-ing & 2 \\
        Accountability & 7 & Private/secret & 2  \\
        Expect/-ed/-ations/-ancy & 6 & Predictability & 2 \\
        Doubt & 6 & Robust & 2 \\
        Open/-ly/-ness & 6 & Responsibility/-ties & 2  \\
        Promise/-d/-s & 5 & Verifiable, verify & 2  \\
        Secure/-ity & 5 & Explainability & 2\\
      
        \bottomrule
    \end{tabular}
    
\end{table}